# PCR-ALERA: A PCR-ELISA analogue but antibody-free DNA detection system

Miles Bonenberger[1], Roland Krämer[1]

[1] Heidelberg University, Inorganic Chemistry Institute, Im Neuenheimer Feld 270, 69120 Heidelberg, Germany.

Correspondence to:

Roland Krämer, phone 0049 6221 548438, fax 0049 6221 548599
E-mail: kraemer@aci.uni-heidelberg.de

**Abstract**

PCR-ALERA is a new concept for DNA detection that combines DNA amplification by polymerase chain reaction (PCR) with an activator-linked enzyme reactivation assay (ALERA). PCR-ALERA resembles the commercial, semi-quantitative PCR-ELISA assay that offers high sensitivity and high-throughput screening in microplates. PCR-ALERA overcomes a major drawback of PCR-ELISA since it does not require the time-consuming handling of expensive antibodies. In the first step of the assay, commercially available modified nucleotides were incorporated during PCR amplification of the target DNA. A selenocysteine containing peptide (the activator) was linked to the amplified DNA by click chemistry. This peptide has been optimized for the highly efficient activation of the latent protease papain by selenolate-triggered disulfide cleavage. The DNA-peptide hybrid was then immoblized in microplate wells and incubated with latent papain. The latter was readily activated and converted a fluorogenic substrate to a fluorescent product with high catalytic turnover. Overall assay time is about 4 hours including PCR.

# 1 Introduction

With the development of the real-time polymerase chain reaction (qPCR) described in 1996 by Heid *et al.*[1] , a new method for quantification and observation of the DNA amplification was achieved. This method allows for a more detailed identification and quantification of DNA compared to standard PCR and is routinely used for diagnostics in clinical microbiology laboratories and many other scientific fields.[2,3] Most recently, the worldwide demand for this technique increased drastically due to the severe acute respiratory syndrome coronavirus 2 (SARS-CoV-2) pandemic. Tailored conditions for the detection of this virus were immediately established and millions of tests have been performed each day.[4] While qPCR is a fast, easily handled, very accurate and relatively low error-prone technique for identification of DNA samples, it is still one of the more expensive detection methods requiring non ubiquitous machinery and expensive chemicals. It is therefore of great interest to develop similar spe-cific but cheaper alternatives which could be utilized in laboratories with standard molecular biological equipment.

Before the invention of PCR, pathogens were mainly identified via the serological method Enzyme-Linked Immunosorbent Assay (ELISA) developed by Engvall *et al.*[5] in 1971. This method allows for a quantitative determination of either antibodies or antigens within a sample by making use of specific interaction between the two. A disadvantage of this method is however the deployment of very specific antibodies which require an elaborate production process often involving animal experiments or long-time cell cultures.[6–8] Nonetheless, this technique is more widely available compared to qPCR with lower costs and good, although not as good as with qPCR, accuracy and is, among others, also applicable for the detection of the coronavirus.[9] The ELISA does however only allow for the detection of proteins or other specific molecules and not actual DNA which is unfavourable for many applications.

In order to bridge this gap, first immunodetections of DNA with biotinylated RNA and monoclonal antibodies were described in 1989 by Coutlée *et al.*[10] and later adapted towards a combination of both PCR and ELISA as first described in 1993 by Albert *et al.*[11]. They used streptavidin coated plates and DNA probes, labelled with biotin at their 5' end, for the immobilization of labelled target DNA which was amplified via PCR in the presence of modified digoxigenin-11-UTP (DIG-dUTP) nucleotides. The nucleotide bound and incorporated digoxigenin was then detected via an enzyme coupled anti-digoxigenin antibody similar to ELISA.

Thus, PCR-ELISA is an immunodetection method which can identify PCR products via enzymatic reaction. This leads to a strong signal amplification which directly reflects the amount of sample immobilized on the microplate via specific DNA-probes, allowing for a comparison between samples in a semi-quantitative manner. When compared to conventional PCR detection via agarose gels, PCR-ELISA features higher specificity, higher sensitivity with lower detection limits, semi-quantifiability rather than a rough estimation of positive and negative signal and a quicker overall process resulting in improved diagnostic times. The method does however not allow for an accurate estimation of absolute DNA amounts which is only possible with qPCR. Furthermore, the initial PCR can be conducted with various primers in order to amplify different sequences within a single sample, identifying various target DNA without cross-reactivity due to very specific hybridization DNA probes.[12–14] This increases the already existent high-throughput ability of the assay manifold, greatly surpassing the detection via agarose or even qPCR in a process that could be easily automated in 96- or even 384-well plates whilst utilizing standard laboratory equipment. An exemplarily direct comparison between qPCR and PCR-ELISA was conducted by Menotti *et al.*[15] who utilized both methods to detect genomic DNA of the parasite *toxoplasma gondii* in clinical patient samples. They found that of 144 clinical symptomatic and with qPCR positively tested patient samples, 15 showed a false negative result with PCR-ELISA. The group of Aini *et al.*[16] published comparable results and described qPCR to be ten times more sensitive compared to PCR-ELISA, which in turn is ten times more specific compared to agarose gel based detection of DNA which was also confirmed by Bagheri *et al.*[17]. Cabrera *et al.*[18] successfully utilized PCR-ELISA for the detection of African trypanosomes in bovine serum and underlined the trade-off with PCR-ELISA concerning maximal sensitivity and availability, stating that the method " […] offers a good combination of simplicity, accuracy, sensitivity and speed." and is especially useful for laboratories with limited equipment. A summarized comparison between qPCR, standard PCR and PCR-ELISA was provided by Sue *et al.*[19] and Tayebeh *et al.*[20] and is depicted in **Table 1**.

**Table 1:** Comparison of the three molecular DNA detection methods: conventional PCR with DNA detection via agarose gel electrophoresis, immunodetection of modified DNA via PCR-ELISA and real-time PCR (qPCR) via fluorescence probes.

| Trait | Conventional PCR | PCR-ELISA | qPCR |
|---|---|---|---|
| **Equipment** | standard laboratory | standard laboratory | special qPCR systems |
| **Reagent cost** | low | moderate | high |
| **Detection limit** | 1 – 10 ng/µL | 0.01 ng/µL | 0.25 pg/µL |
| **Quantitative ability** | none | semi-quantitative | quantitative |
| **Toxicity** | methylene blue (low)<br>ethidium bromide (high) | none | none |

By now, PCR-ELISA has been successfully utilized in a wide variety of medical diagnosis, such as in clinical cancer research [21–24], fungal infections of susceptible patients [25–27] as well as detection of viruses and bacteria like HIV [17], Hepatitis [28,29], Epstein-Bar [30], Mycobacterium tuberculosis [31] and Methicillin-resistant Staphylococcus aureus (MRSA) [32]. Furthermore, the method provided a cheap and efficient tool for genotyping of blood groups [33] and successfully identified point mutations connected to the blood ailment beta thalassemias [34], thus substituting very expensive and protracted DNA sequencing. PCR-ELISA was also used for parasite tracings like Trypanosomes [35] and Leishmania infantum [36,37] among others. Another important sector of PCR-ELISA application is the food industry, where a vast number of foodborne pathogens like Brucella [38], Escherichia coli [39] and Salmonella [40], to mention only a few, have been efficiently detected. Besides pathogens, PCR-ELISA has been deployed to detect and identify genetically modified organisms in food.[41]

Due to the similar process based on traditional ELISA, PCR-ELISA has the previously discussed disadvantageous need for antibodies which are not only costly but have to be produced in an elaborate process, often involving animal experiments. Furthermore, the antibodies have to be coupled to the utilized enzyme, rendering the whole process non-autonomous for the standard laboratory, although progress has been made towards facilitation of the production of improved PCR-ELISA specific antibodies.[42]

In order to eliminate these disadvantages while maintaining the efficiency and specificity of the PCR-ELISA, a completely new method of DNA detection without the need for antibodies would be desirable. It would moreover be ideal if the method would be self-sustaining for standard laboratories.

This herein designed and so-called PCR-ALERA (PCR-activator linked enzyme reactivation assay) involves enzymatic signal amplification similar to ELISA, whereas the unconjugated enzyme is applied in a reversible inhibited form which is reactivated in solution by an activator linked to amplified target DNA. The integral steps and tools of the process are depicted in **Figure 1**.

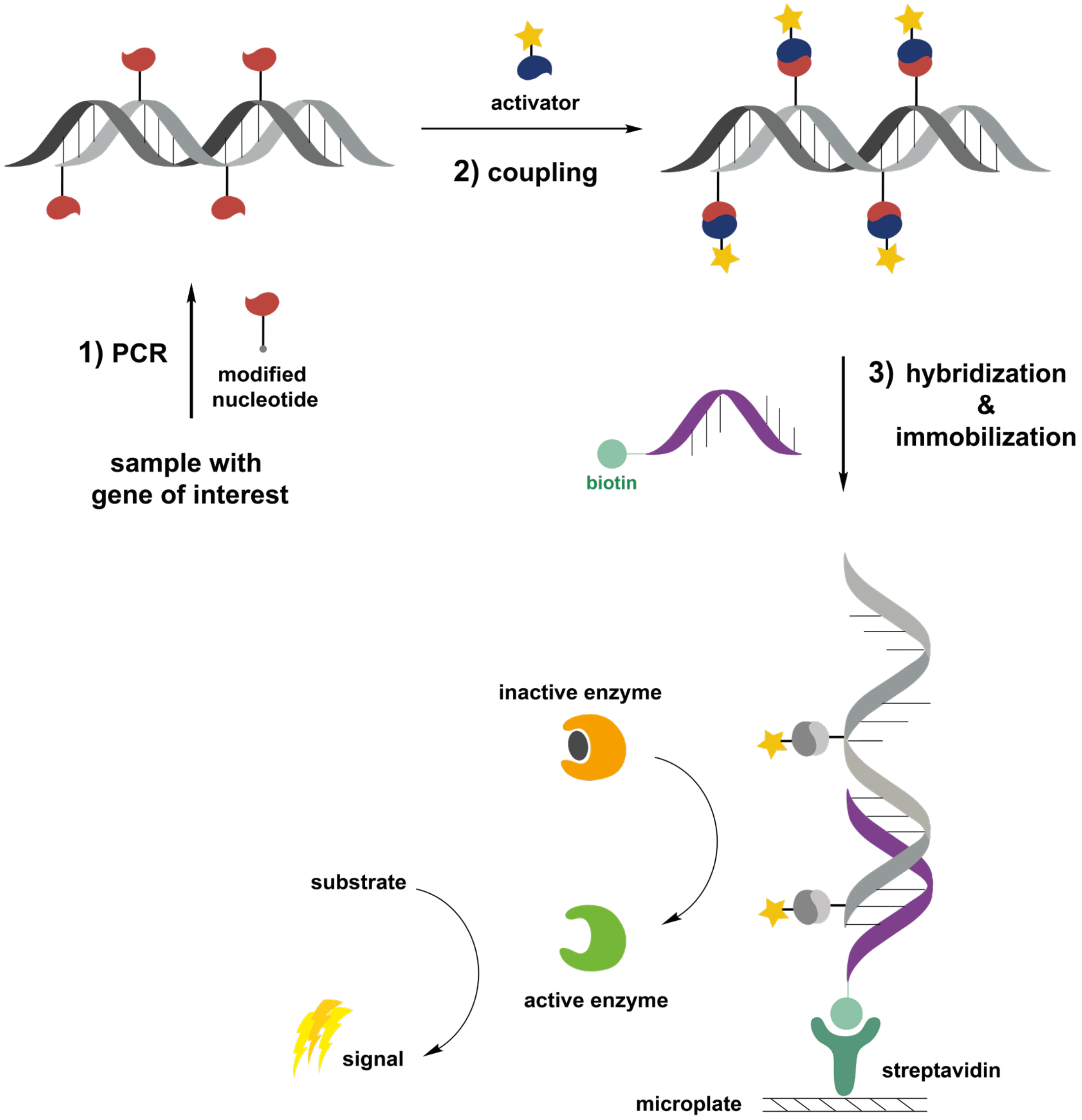

**Figure 1:** Principle of the indirect PCR-ALERA. 1) Gene of interest is amplified via PCR leading to incorporation of modified nucleotides. 2) Reactive moieties of modified and incorporated nucleotides are coupled to corresponding moieties in enzyme activator molecules. 3) Target DNA with linked enzyme activator hybridizes with specific biotinylated DNA probes and are immobilized on streptavidin coated plates in the process. Immobilized DNA with linked enzyme activator reactivates inhibited enzyme which in turn processes the substrate leading to an amplified signal.

Amplification of target DNA is achieved by PCR during which modified nucleotides are incorporated. The modified nucleotides contain a dibenzocyclooctyne (DBCO) moiety which is employed in the next step in a ring-strain promoted click reaction between the DNA amplificate and an azide containing activator peptide. The synthesized peptide contains selenocysteine which in turn functions as an activator for the utilized enzyme papain. In the so-called indirect PCR-ALERA, the DNA-peptide-hybrid is hybridized to a complementary biotinylated probe DNA and subsequently immobilized on a streptavidin coated plate in the process. The direct PCR-ALERA omits the biotinylated probe and utilized biotinylated primers to directly tag the amplified DNA. The cysteine protease papain was reversibly inhibited by reaction with methyl methanethiosulfonate beforehand. The peptide which is coupled to the amplified DNA promotes a selenol/disulfide exchange reaction, reactivating papain in the process. Active papain processes its substrate Z-Phe-Arg-AMC, leading to an increase in fluorescence which can be detected by most conventional plate readers. A direct comparison between the traditional PCR-ELISA and the herein developed PCR-ALERA is depicted in **Figure 2**.

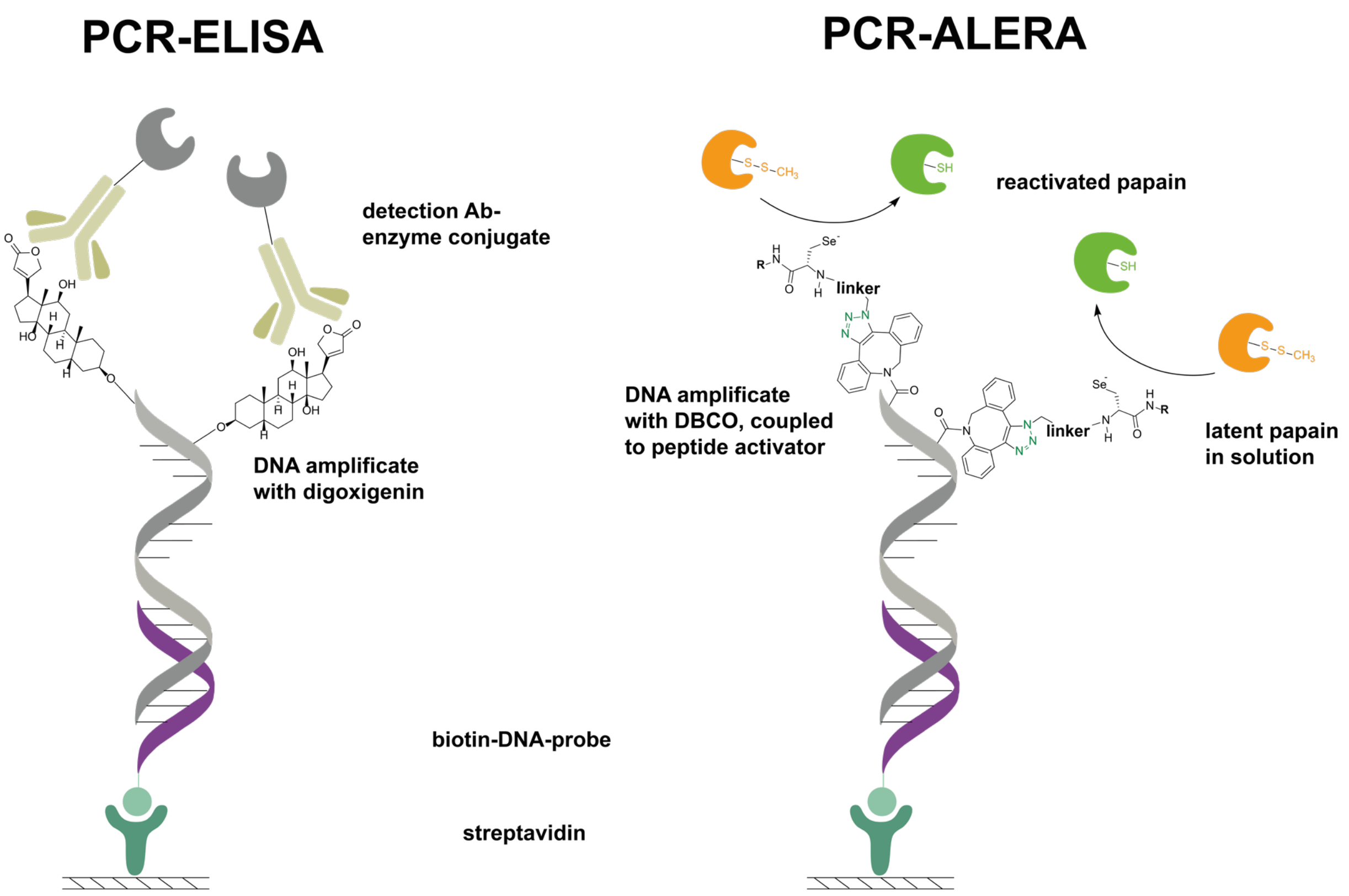

**Figure 2:** Comparison between the final setups of the traditional PCR-ELISA and the PCR-ALERA. Both methods utilize biotinylated probe DNA for hybridization and immobilization of target DNA. PCR-ELISA is based on digoxigenin labelled DNA which is detected via antibodies. Antibody conjugated enzyme processes substrate producing a detectable signal. The PCR-ALERA is based on DBCO labelled DNA which reacts with an activator peptide in a ring-strain promoted click reaction. The conjugated peptide activates latent papain in solution which in turn processes substrates, leading to a fluorescence signal.

As depicted, the PCR-ALERA eliminates the need for antibodies and utilizes a specifically designed activator for the precise activation of latent papain, leading to substrate processing in solution rather than at the solid-solution interface.

# 2 Results and discussion

## 2.1 Papain Activators:

In order to develop a sensitive PCR-ALERA, finding an efficient papain-activator was imperative. In 1966 it was proven by Sluyterman *et al.*[43] that the cysteine protease papain is recovered partly as an inactivated mixed disulfide between $Cys_{25}$ of the active site and a free cysteine, which can be completely restored through thiol activators. This means that papain is only available in an already reversible inactive form but the inactivation is not complete and thus papain shows a residual activity of up to 50% when freshly prepared.[44] Furthermore, an efficient way of complete and reversible deactivation of papain via thiosulfonates was described by Shaked *et al.*[45]. This led to papain becoming an important part in the family of model proteins for the examination of disulfide exchange reactions and development of different most efficient disulfide reducing agents[46–49]. Utilizing this system, a series of candidates were screened in our group, comprising aromatic thiols, dithiols, aliphatic thiols as well as cysteine containing peptides of different sequences. It was shown that thiophenol derivatives were the most efficient activators followed by cysteine containing peptides (results not shown). Due to many disadvantages of thiophenol derivatives (poor solubility, toxicity, volatility and more), it was decided to pursue activators comprised of small peptides. Screening of more than 20 different peptides of different lengths (between 2 and 6 amino acids) procured the tripeptide $H_2N$-CLR-$CONH_2$ to be the most efficient of the tested peptide activators. Due to the differences in properties between sulfur and selenium, as discussed in great detail by Wessjohann *et al.*[50], reaction rates in thiol/disulfide-like exchange reactions utilizing selenols and thiols differ greatly as shown by different groups[51–53] and conveniently summarized by Nauser *et al.*[54]. Expecting to yield an even more efficient papain activator, cysteine was thus substituted with its selenol analogue selenocysteine in the previously identified potent activator $H_2N$-CLR-$CONH_2$. The synthesized tripeptide $H_2N$-ULR-$CONH_2$ (the one letter code for selenocysteine is U) was subsequently screened and identified as an extraordinary papain activator (results not shown).

Before DNA-detection experiments via PCR-ALERA could be performed, it was important to build a model compound comprising a nucleoside or nucleotide coupled to the efficient papain activator and subsequently test the ability to reactivate papain in order to evaluate the results under more simple assay conditions. To this end, a spacer was incorporated (see Figure 3) in order to attach a functional group yielding azido-$PEG_3$-ULR, which is able to interact with the target molecule on the modified nu-cleotide or nucleoside.

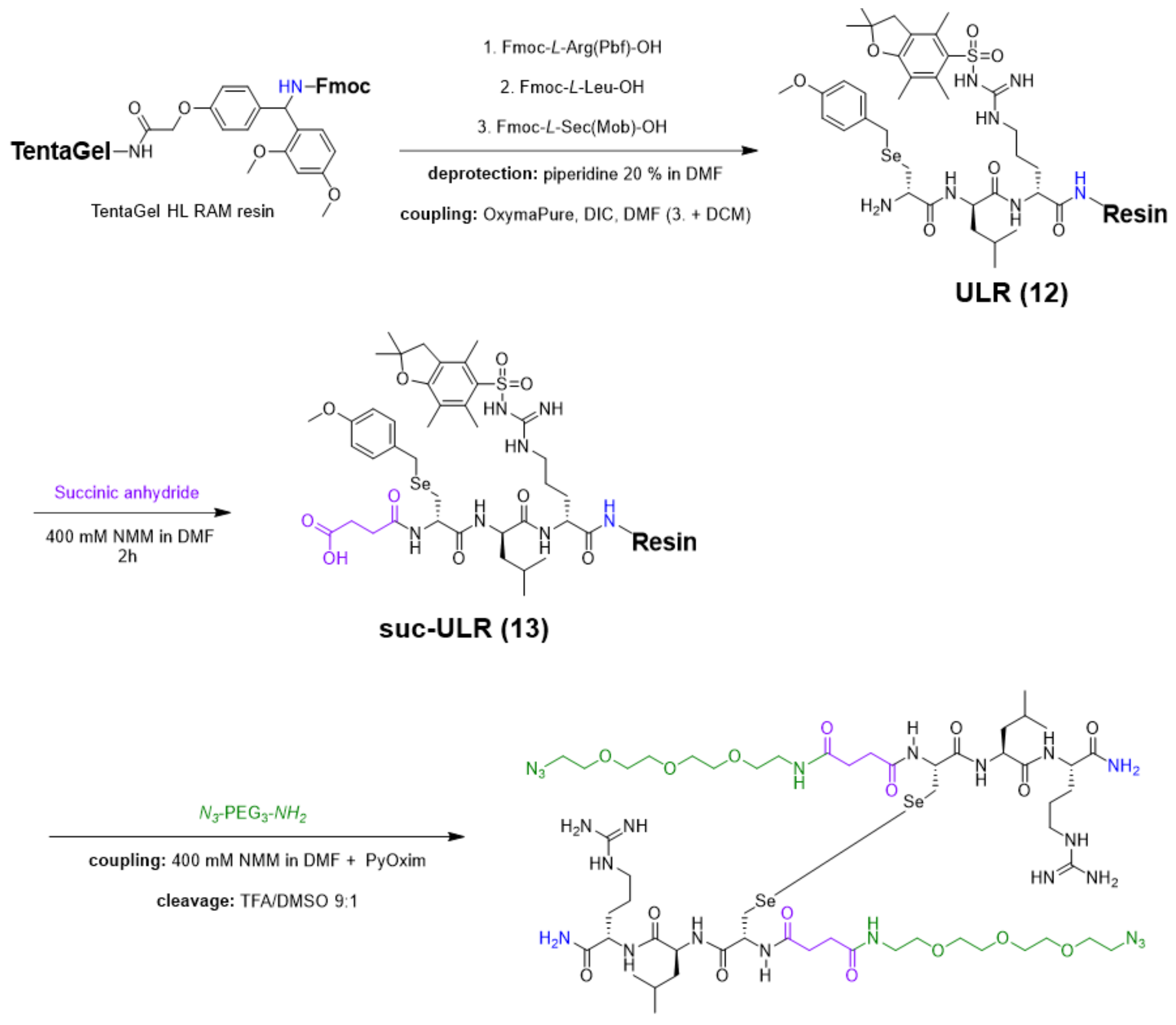

Azido-PEG$_3$-ULR dimer (15)

**Figure 3:** Synthesis scheme of the solid phase peptide synthesis towards azido-PEG$_3$-ULR (**15**).

The synthesis of the modified deoxyuridine 18 (see figure 4) was adapted from Jäger *et al.*[55] with coinciding analytical data (not shown). The nucleoside-peptide-hybrid model compound dU-triazole-PEG$_3$-ULR (21) was consequently synthesized in two steps, first by a standard amine-carboxyl coupling between (E)-18 and commercially available dibenzocyclooctyne-amine and secondly by coupling 20 with the previously synthesized peptide dimer azido-PEG$_3$-ULR (15) with TBTU, HOBt and DIPEA. The subsequent copper-free ring-strain promoted click reaction progressed readily in acetonitrile/water to yield 64% dimeric nucleoside-peptide-hybrid after purification via preparative HPLC. The formation of the triazole leads to two regioisomers, a 1,4-triazole and a 1,5-triazole. With two coupled DBCO nucleosides, this leads to a total of three possible isomers of dU-triazole-PEG$_3$-ULR (21) which were however not separated due to their assumed identical reactivity.

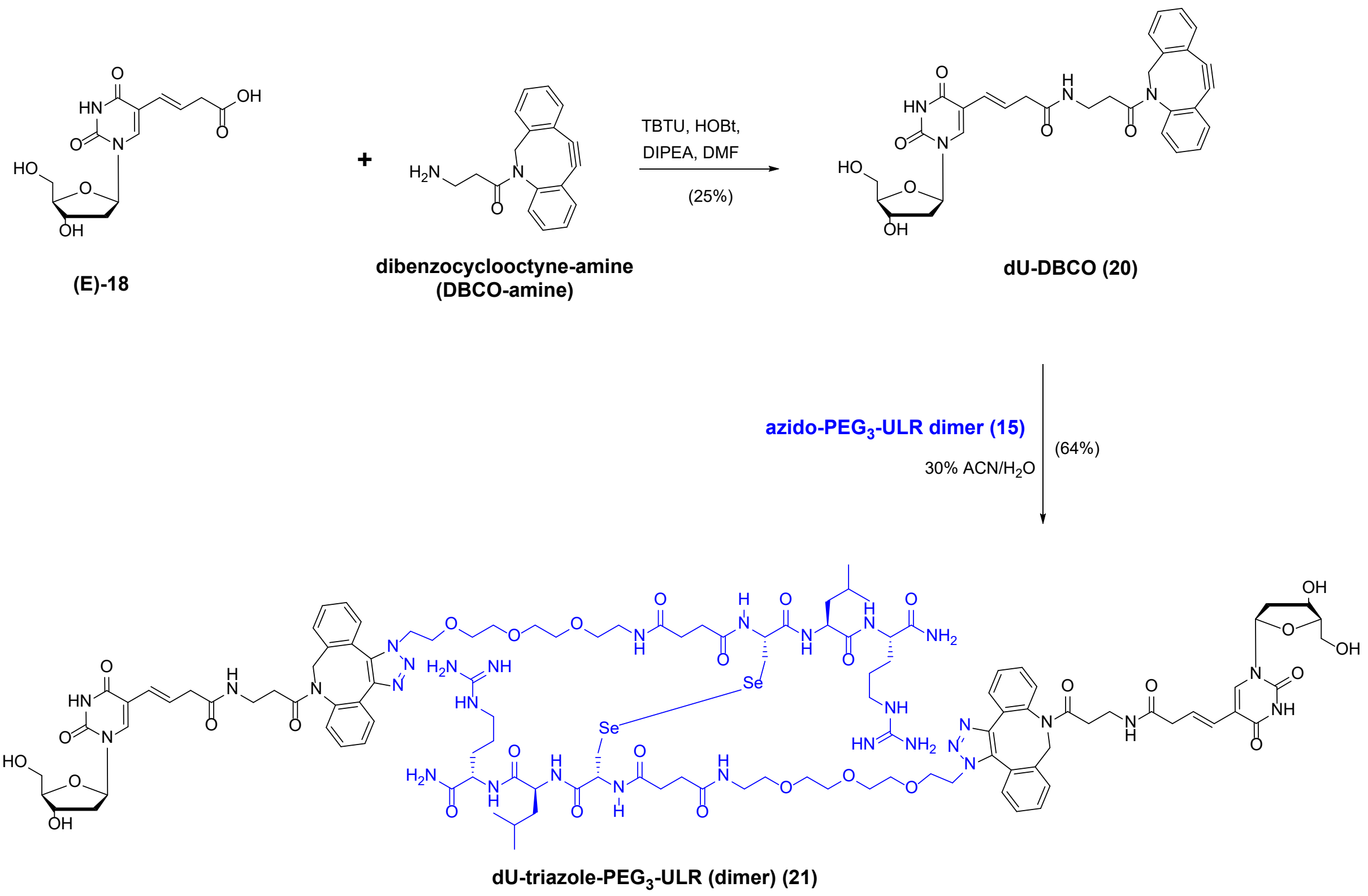

**Figure 4:** Reaction pathway towards the model compound dU-triazole-PEG$_3$-ULR (**21**). Synthesis was carried out in two steps by coupling **(E)-18** and DBCO-amine and subsequent ring-strain promoted click reaction with previously synthesized peptide dimer azido-PEG$_3$-ULR (**15**).

Every peptide intermediate of the multi-step synthesis was analysed and utilized individually in a papain reactivation assay in order to get a clear overview of papain preferences of the synthesized molecules (results not shown). It was found that the final model compound 21 increased the papain reactivation manifold over the simple tripeptide *$H_2N$*-ULR-*$CONH_2$* (results not shown).

Since selenols readily oxidize to diselenides at physiological conditions in the presence of molecular oxygen, the diselenide activators have to be reduced prior to papain activation. This can be achieved via addition of dithiols (most monothiols lack the necessary redox potential for reduction of diselenides as described by Singh *et al.*[56], unless applied in 1000-fold excess [57]) as first reported by Dickson *et al.*[57] in 1969 and since wildly used by other groups. Those dithiols are however fairly potent papain activators themselves, rendering them unsuitable for the reduction of utilized diselenide peptides in the assay. Commercially available TCEP on the other hand is a most efficient disulfide reducing agent with many beneficial properties as described by Winther *et al.* [58] and was further proven to be a poor papain activator in the required low concentrations (results not shown) making it an ideal reducing agent in a one-pot assay.

The model compound dU-triazole-PEG$_3$-ULR (**21**) proved to be the most efficient papain activator tested in our group to date and was therefore compared to other compounds described as potent reducing agents in literature.[45,59,60] To this end, cysteine ethyl ester, DTT, DTBA (**9**) and SeDTBA (**11**) were chosen for a direct comparison. DTBA (**9**) as well as its selenium analogue was synthesized and identified according to literature.[59,61] Other potent reducing agents like BMS, DMH and DTA presented by the Whitesides group[46,47] were not investigated further due to the lack of functional groups necessary for attachment to a linker structure and ultimately towards a model compound and the DNA itself. The assay was conducted with 1 nM latent papain in well buffer (0.1 mM phosphate buffer pH 7, 4 mM EDTA, 0.01% Brij35), 2 µM Z-Phe-Arg-AMC substrate, 1 µM TCEP and with the activators in three activator concentration series from 50 pM – 750 pM, 1 nM – 15 nM and 10 nM – 10 µM. The resulting reaction rates were normalized against the background activity of latent papain without addition of activators and the resulting normalized rates are presented in **Figure 5** for the concentration range of 50 pM – 100 nM.

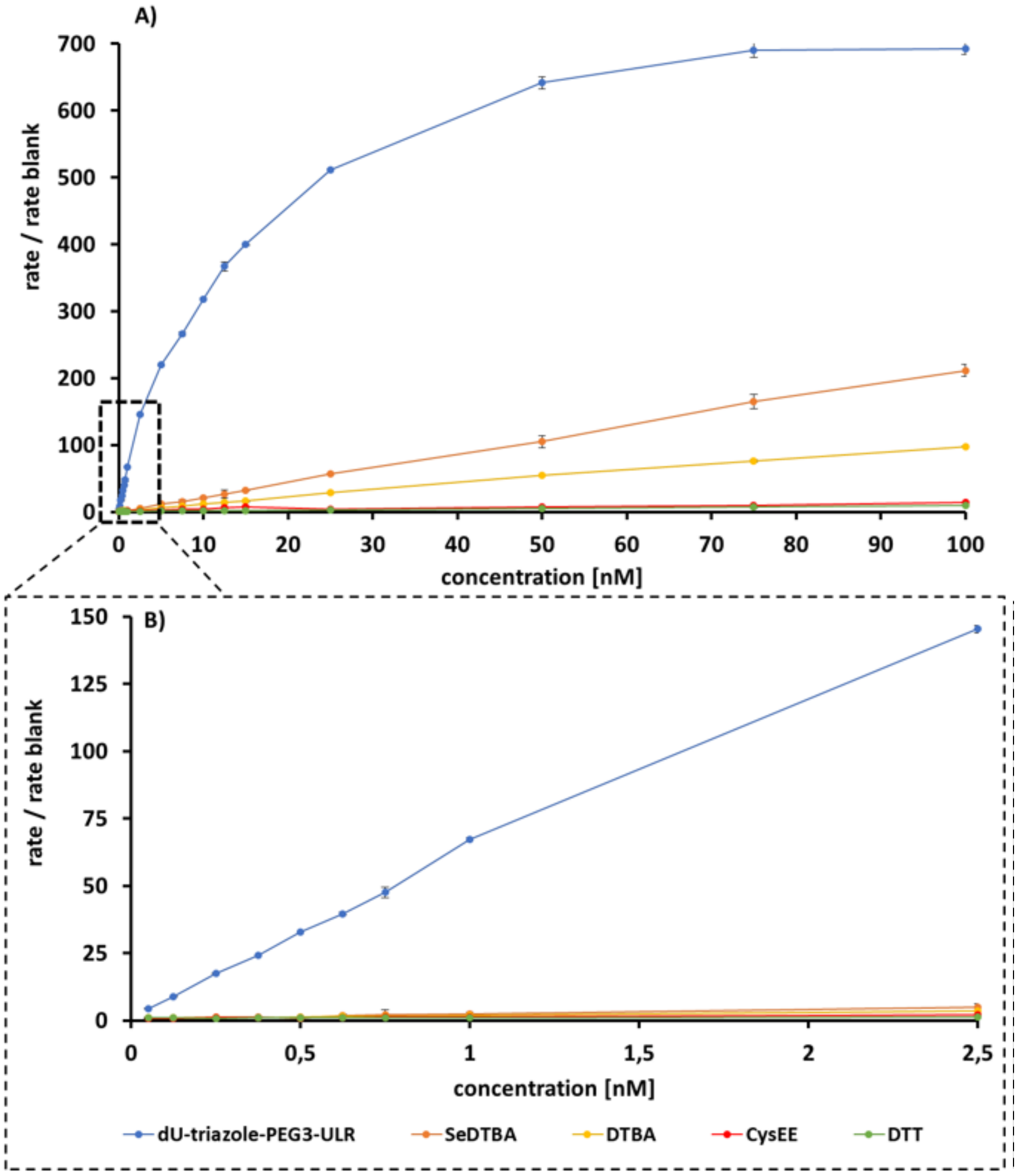

**Figure 5:** Normalized ratios of rate/rate blank of substrate hydrolysis (2 µM) after reactivation of latent papain (1 nM) with the displayed activators at concentrations of 50 pM – 100 nM in well buffer (0.1 mM phosphate buffer pH 7, 4 mM EDTA, 0.01% Brij35) in the presence of 1 µM TCEP. Measurements were performed in duplicate and standard deviations are shown. Blank represents the residual background activity of papain without addition of activator. Measuring points were connected for better illustration. The abbreviation "CysEE" stands for *L*-cysteine ethyl ester. A) Overview of reaction rates of activators in the concentration range of 50 pM – 100 nM. B) Zoom-in of the concentration range between 50 pM and 2.5 nM.

As seen in **Figure 5** A, the reaction rates after papain activation via the model compound dU-tria-zole-$PEG_3$-ULR (**21**) are considerably surpassing those of all others. SeDTBA (**11**) is the second most efficient activator with rates almost twice as high compared to DTBA (**9**). Cysteine ethyl ester and DTT produce no substantial rate increase in this low concentration range and are viable only in concentra-tions above 1 µM (results not shown). Even at low picomolar concentrations the partial reactivation of latent papain with dU-triazole-$PEG_3$-ULR (**21**) is efficient enough to produce a fluorescence distinguish-able from the background activity.

Lukesh *et al.*[59] reported a reactivation rate of latent papain with DTBA (**9**) that is 14 times higher com-pared to activation with DTT. This claim is corroborated by the findings in this work, where DTBA (**9**) produced rates 13.6 times higher than DTT at a concentration of 50 nM. The model compound dU-tri-azole-$PEG_3$-ULR (**21**) is however strongly exceeding the potency of DTBA (**9**) at every concentration. At 50 nM the model compound produces rates 128 times higher than DTT and 9 times higher than DTBA (**9**). At a concentration of 10 nM the difference increases further and the rates are 159 times higher than DTT and 23 times higher than DTBA (**9**). With decreasing concentration, the potential of dU-tria-zole-$PEG_3$-ULR (**21**) becomes even more apparent because all other activators fail to activate latent papain to a significant degree. Below the concentration of 2.5 nM none of the other activators produce rates higher than four times the background activity while the model compound shows such a signifi-cant result at a concentration as low as 50 pM. At a half equimolar concentration of papain (500 pM) the activator still produces a rate 33 times above the background activity.

To the authors best knowledge, no other in literature reported papain activator shows comparable efficiencies. It was therefore concluded that the peptide activator azido-$PEG_3$-ULR (**15**) should be suit-able for attachment to commercially purchased DBCO-modified dUTP nucleotides towards DNA-de-tection experiments.

## 2.2 Direct PCR-ALERA:

Since the model compound showed promising results, the peptide intermediate azido-$PEG_3$-ULR (**15**) was utilized in DNA-detection assays. Furthermore, a hetero peptide dimer was synthesized consisting of two different monomeric peptides, one identical to azido-$PEG_3$-ULR (**15**) with an azide moiety and a second one of similar structure containing an amine moiety instead, in order to prevent possible cross-linking of modified nucleotides within the DNA. Both peptides were utilized and compared in the fol-lowing DNA-detection assays and the structure of both peptides as well as the commercially available dUTP-DBCO nucleotide are shown in figure 6.

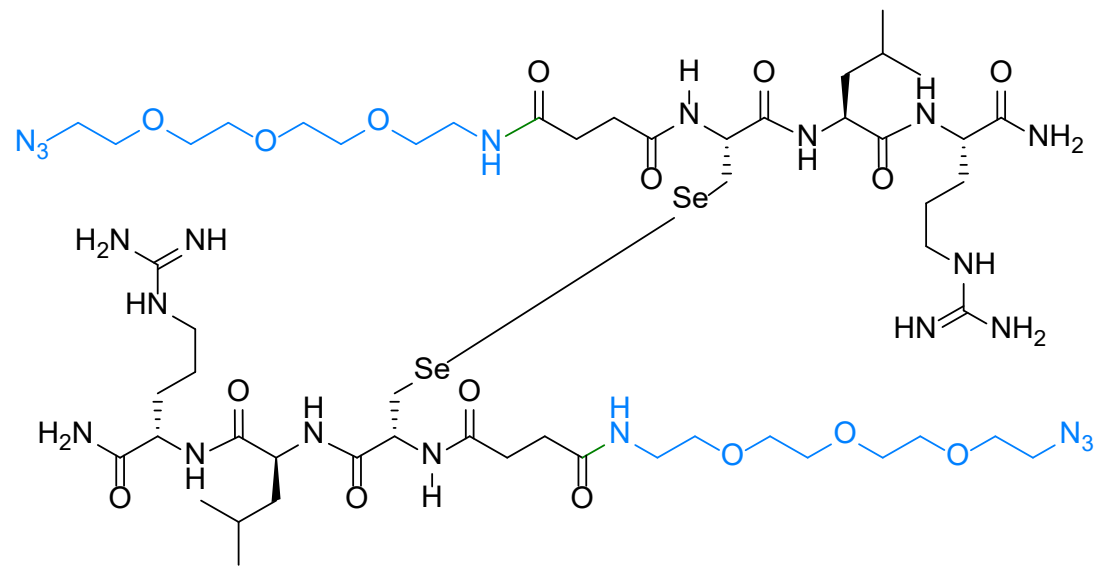

**azido-PEG$_3$-ULR dimer (15)**

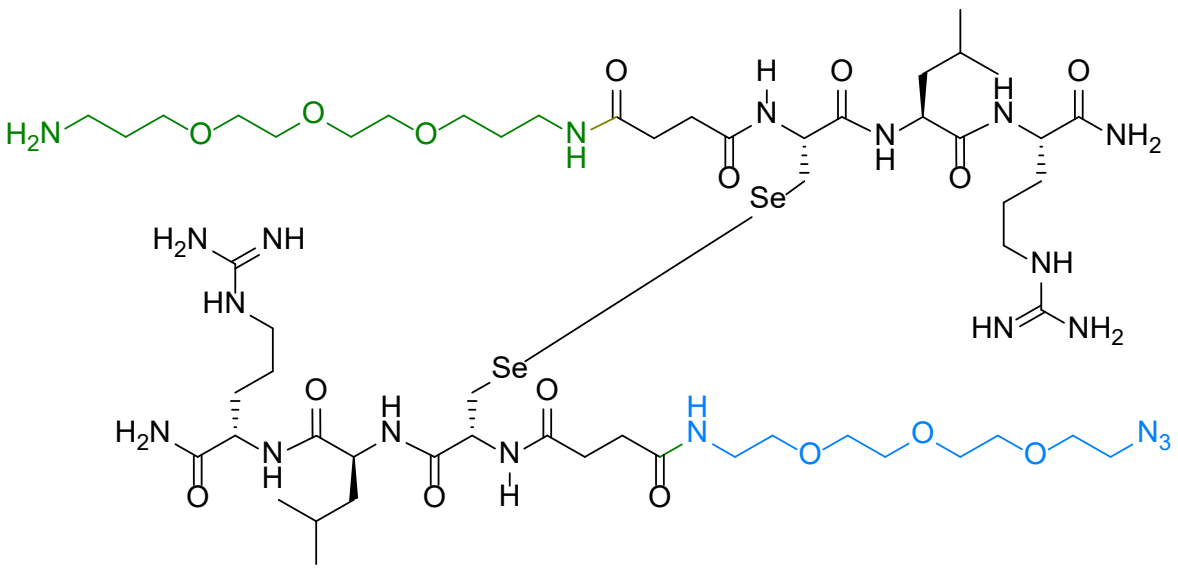

**azido-amino-PEG$_3$-ULR dimer (16)**

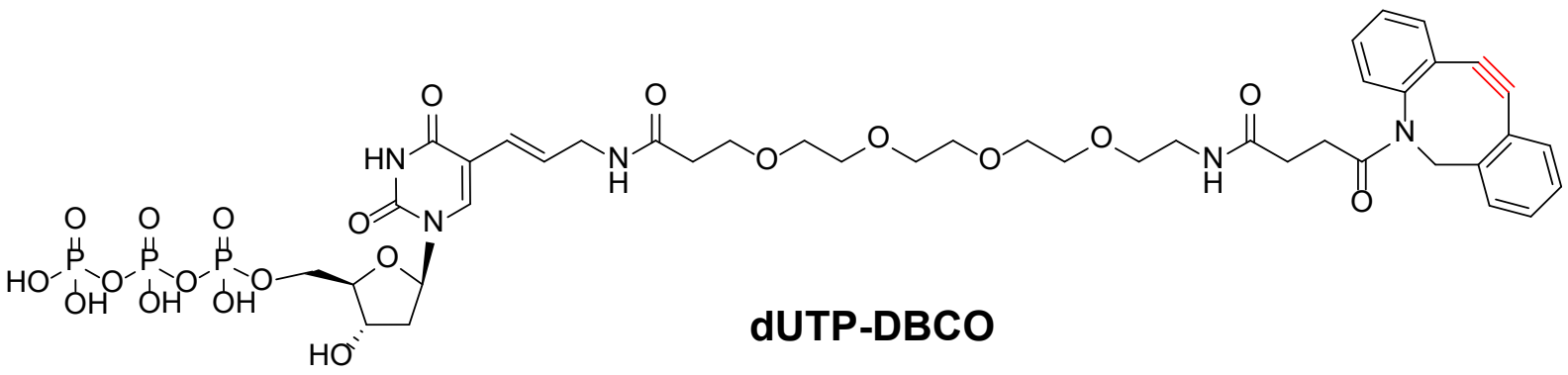

**dUTP-DBCO**

**Figure 6:** Structure of the utilized peptide-dimers azido-PEG$_3$-ULR (**15**) and azido-amino-PEG$_3$-ULR (**16**) as well as the commercially available dUTP-DBCO nucleotide derivates. The azide linker is marked in blue and the amine linker is coloured in green. The alkyne moiety, marked in red, reacts with the azide moiety to form a triazole-ring in a copper-free ring-strain promoted click reaction.

SARS-CoV-2-S DNA (in the following called Spike DNA) was amplified via PCR and DBCO modified dUTP nucleotides were incorporated in the process. The PCR reaction mixture contained two different amounts of dUTP-DBCO. Amplification of a negative control DNA (subsequently referred to as DNA-) was achieved with 100% dTTP and 0% dUTP-DBCO content in the reaction mixture. DNA amplification of a positive control DNA (subsequently referred to as DNA+) contained 75% dUTP-DBCO and 25% dTTP nucleotides. A detailed overview of the process for the DNA-detection assay, in this case the previously described direct PCR-ALERA (without hybridization probe) of biotinylated DNA is depicted in **Figure 7**.

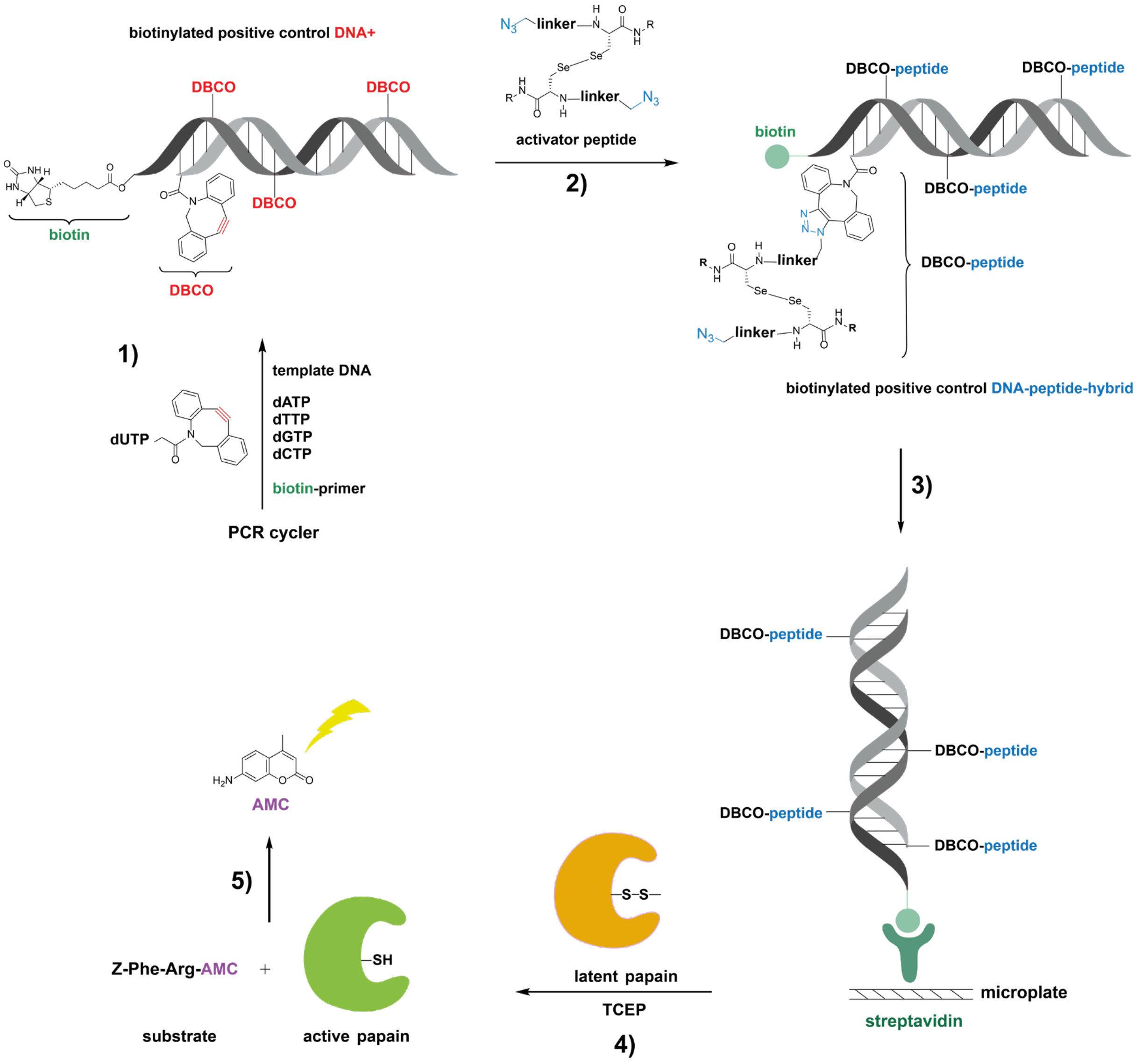

**Figure 7:** Overview of the direct PCR-ALERA with a biotinylated positive control DNA. 1) PCR with template DNA, standard nucleotides, biotinylated primers, Taq polymerase and modified dUTP-DBCO nucleotides produce a DNA+ amplificate which is biotinylated and contains DBCO moieties. 2) The PCR mixture is incubated with activator peptide containing an azide moiety for the ring-strain promoted click reaction. 3) The raw (or purified) biotinylated DNA-peptide-hybrid (DNA+ph) mixture is transferred to the streptavidin coated microplate and subsequently immobilized. 4) Latent papain, TCEP and substrate is added to the wells. 5) Activated papain processes the substrate and nascent free AMC is detected via a fluorometer.

### 2.2.1 Single-step DNA purification via Sephadex

The amplified DNA is incubated with activator peptide before being transferred into streptavidin coated wells. It is therefore advisable to include a purification step in order to remove excess of unreacted free activator and other PCR compounds. G-50 Sephadex Quick Spin columns were chosen because they efficiently separate modified DNA from excess peptide, primer and nucleotides in a fast and simple process. Biotinylated Spike DNA+ and DNA- were prepared by PCR and then divided into two equal portions of 50 µL each. The raw PCR solution was subsequently incubated with equimolar amounts of either azido-PEG$_3$-ULR (**15**) or azido-amino-PEG$_3$-ULR (**16**). After incubation, all four DNA portions were purified via sephadex columns and two different dilutions of each DNA solution (1:100 and 1:50 in well buffer) were prepared. 300 µL of each solution was added to the streptavidin coated Greiner-Bio-One plate. After immobilization and washing steps, the reactivation assay was conducted with latent papain as described previously. The resulting reaction rates were normalized against the rates of the corresponding negative control DNA-. These normalized rates are shown in **Figure 8**. Meas-urements were performed in triplicate and standard deviations are shown.

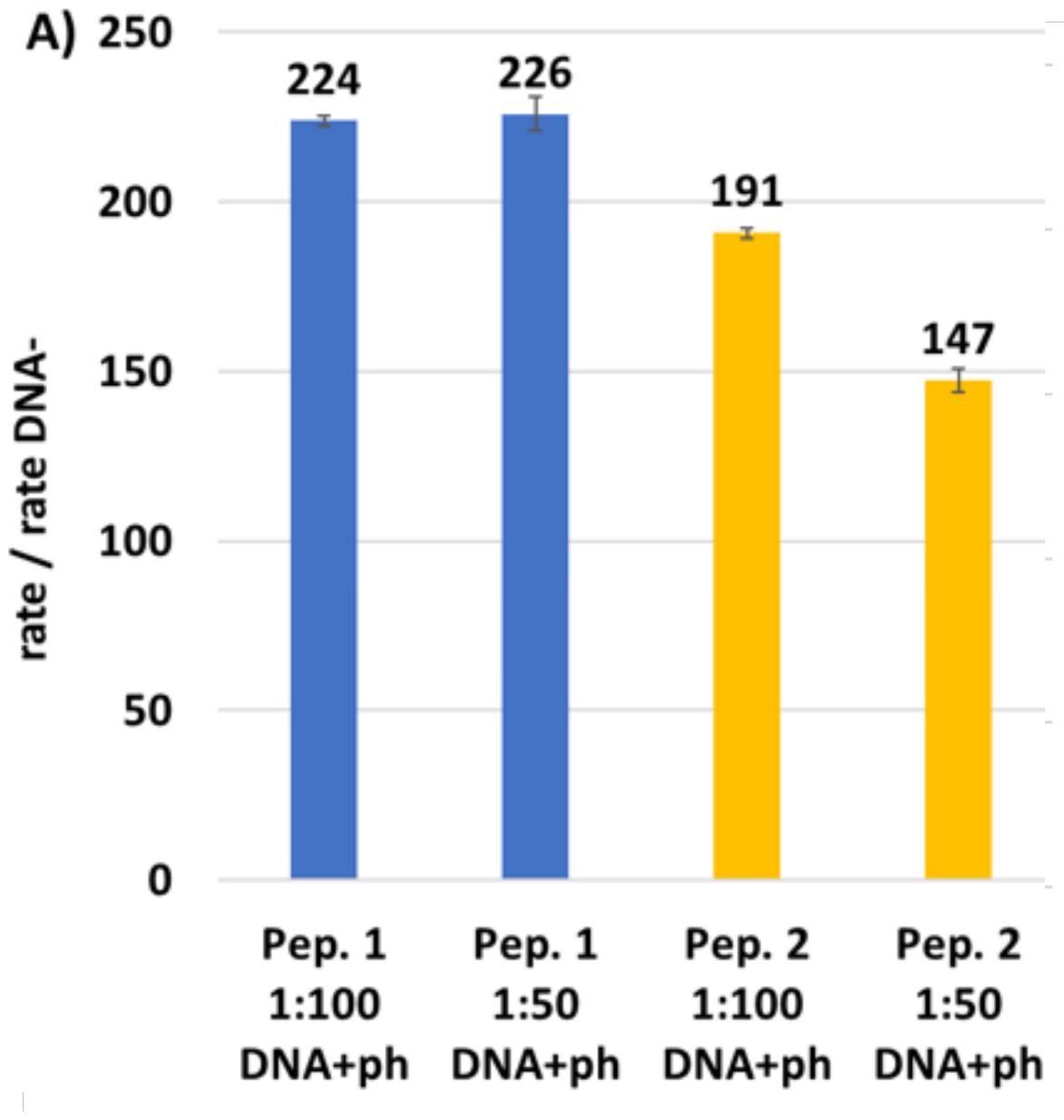

**Figure 8:** latent papain (1 nM) reactivation assay with immobilized DNA-peptide-hybrid (DNA+ph) in well buffer (0.1 mM phosphate buffer pH 7, 4 mM EDTA, 0.01% Brij35) in the presence of 1 µM TCEP and 2 µM substrate. A) Normalized ratios of rate positive control (DNA+ph)/rate negative control (DNA-). Measurements were performed in triplicate and standard deviations are shown. The utilized peptides in this assay were azido-PEG$_3$-ULR (**15**) (Pep. 1) and azido-amino-PEG$_3$-ULR (**16**) (Pep. 2). 1:50 and 1:100 refers to the dilution factor of DNA solution that was added to each well after purification.

As seen in **Figure 8** doubling of DNA-peptide-hybrid per well results only in an insignificant increase in reaction rates in both cases. Further diluting DNA samples will resolve this (see figure 10+11). This is possibly due to a maximal biotin binding capacity of the wells. Normalized rates of positive controls incubated with azido-PEG$_3$-ULR (**15**) are slightly higher compared to rates reached with azido-amino-PEG$_3$-ULR (**16**). Even the lowest activation rate with azido-amino-PEG$_3$-ULR (**16**) is 147 times higher compared to the negative control, whereas with azido-PEG$_3$-ULR (**15**) rates 220 times higher are achieved. These values prove, that the single-step purification is suitable for the assay procedure and yield pure samples.

The described execution of assay steps of the direct PCR-ALERA result in a very sensitive DNA-detection system that can differentiate positive and negative control DNA by over factor 200.

### 2.2.2 Omitting preceding DNA purification

Although the purification with sephadex columns is quick, easy and efficient, omitting purification al-together would accelerate the process even more and abolish the need for specific additional chemi-cals and consumables. It was therefore tested if the purification step can be omitted.

To this end, biotinylated Spike DNA+ and DNA- were prepared by PCR and subsequently incubated with equimolar amounts of azido-amino-PEG$_3$-ULR (**16**). Three different dilutions of each DNA solution were prepared (1:600, 1:100 and 1:30 in well buffer) and 300 µL of each dilution was added to the streptavidin coated plate in triplicate. After immobilization and washing steps, the reactivation assay was conducted as previously described. The reaction rates were normalized against the rates of the corresponding negative control DNA-. These normalized rates are shown in **Figure 9**. Measurements were performed in triplicate and standard deviations are shown.

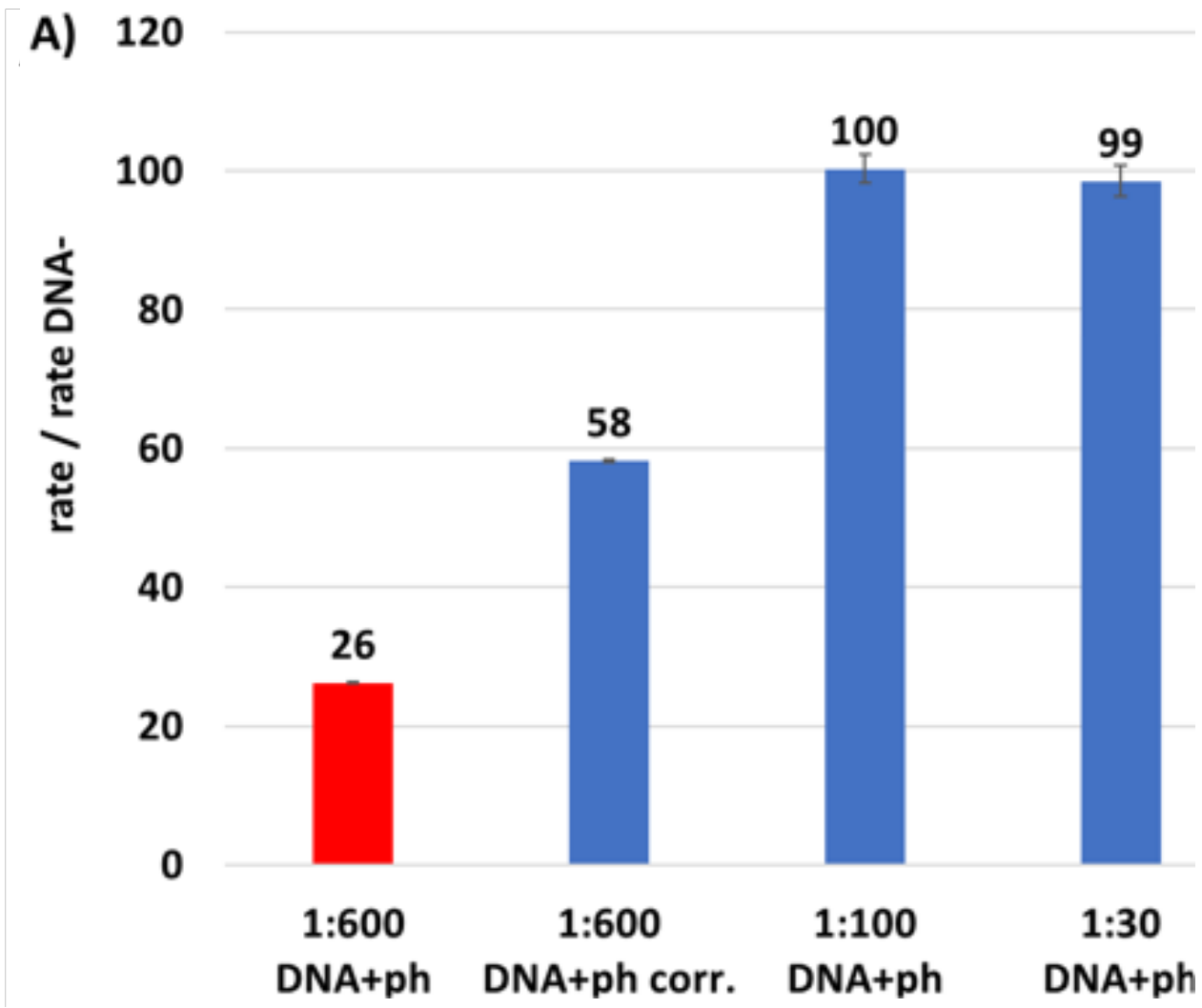

**Figure 9:** Latent papain (1 nM) reactivation assay with immobilized DNA-peptide-hybrid in well buffer (0.1 mM phosphate buffer pH 7, 4 mM EDTA, 0.01% Brij35) in the presence of 1 µM TCEP and 2 µM substrate. The utilized peptide in this assay was azido-amino-PEG$_3$-ULR (**16**). A) Normalized rates of positive control (DNA+ph) against rate of negative control (DNA-). Measurements were performed in triplicate and standard deviations are shown. 1:600, 1:100 and 1:30 refers to the dilution factor of the reaction mixture that was added to each well after the click reaction. The rate of 1:600 DNA- in B (marked in red) shows unnaturally high values compared to the higher concentrated negative controls and thus the corresponding normalized rate of 1:600 DNA+ in A (marked in red) is very high. In order to supply a useful estimation, an additional corrected (corr.) rate is supplied by division of the rate of 1:600 DNA+ph through the rate of 1:100 DNA- instead.

As seen in **Figure 9**, the normalized rate of 1:600 DNA+ph (marked in red) is approximately 50% lower compared to the rate with 1:100 DNA+ph. Normalized rates of 1:100 and 1:30 DNA+ph are almost identical. The observed normalized rates are generally lower by factor 2 compared to the last experiment. This most likely means that residual peptide has not been completely removed from the wells after several washing steps and unspecifically binds to the wells, if no preceding purification step is introduced. Furthermore, the small differences between dilutions of 1:100 to 1:30 DNA+ shows that, as previously described, the highest DNA+ph concentration most likely reaches the maximal binding capacity of the wells.

Considering all factors, the assay without purification, despite its shortcomings, still allows for a very sensitive differentiation of negative and positive control DNA by factor 50 – 100. Overall assay time including PCR amplification is about 4 hours. Further concentration optimizations could possibly allow for a sensitivity close to the previously presented single-step purification assay.

### 2.2.3 Comparison between biotinylated and non-biotinylated positive control DNA

In the following experiment it was examined if biotinylated positive control DNA+ph and a non-biotinylated positive control DNA+ph can be differentiated and if the non-biotinylated DNA+ph could possibly affect papain activation. To this end, biotinylated Spike DNA+ and non-biotinylated Spike DNA+ were prepared by PCR. Both PCR solutions were subsequently incubated with equimolar amounts of azido-amino-PEG$_3$-ULR (**16**). Both DNA solutions were then purified via sephadex column and the eluates were diluted 1:600, 1:100 and 1:30 in well buffer and 300 µL of each solution was added to the streptavidin coated plate in triplicate. After incubation and washing steps, the reactivation assay was prepared with latent papain as previously described. The resulting rates of the biotinylated DNA+ph were normalized against the rates of the corresponding non-biotinylated DNA+ph (see **Figure 10** A) as well as against the rate of residual active papain without addition of DNA or peptide (blank) as seen in **Figure 10** B.

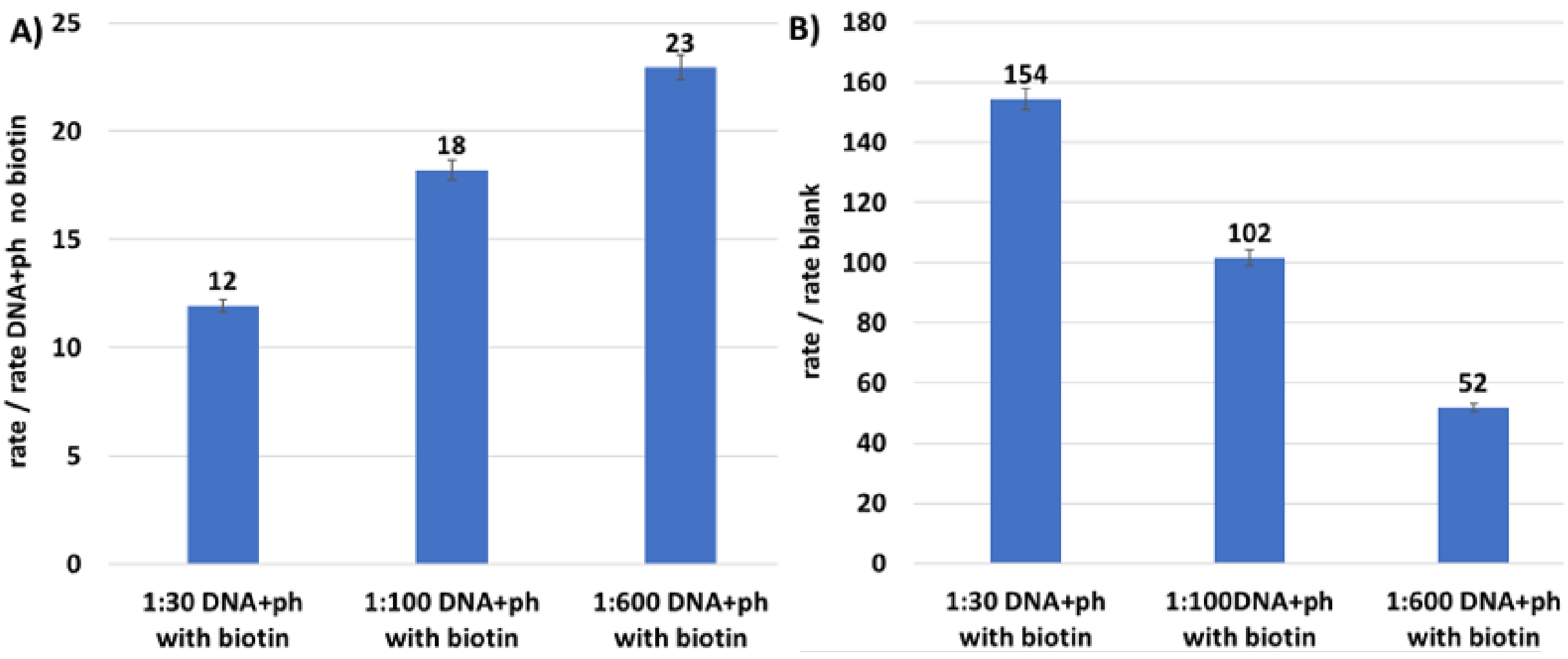

**Figure 10:** Latent papain (1 nM) reactivation assay with Spike DNA-peptide-hybrid in well buffer (0.1 mM phosphate buffer pH 7, 4 mM EDTA, 0.01% Brij35) in the presence of 1 µM TCEP and 2 µM substrate. A) Normalized ratios of rate biotinylated positive control (DNA+ with biotin)/rate non-biotinylated positive control (DNA+ no biotin). B) Normalized rates of biotinylated positive control (DNA+)/rate of residual active papain (blank). Measurements were performed in triplicate and standard deviations are shown. The utilized peptide in this assay was azido-amino-PEG$_3$-ULR (**16**). 1:30, 1:100 and 1:600 refers to the dilution factor of DNA solution that was added to each well after the click reaction and subsequent purification.

The results show, that the normalized rates against those of the corresponding non-biotinylated DNA+ph are rising with decreasing DNA+ph concentrations as seen in **Figure 10** A whereas normalized rates with the blank rate greatly increase with increasing DNA+ph amounts. This means that the non-biotinylated DNA+ph forms unspecific interactions with the well surface in a similar manner as observed for the free activator in the previous experiment. Unlike residual peptide, non-biotinylated DNA-peptide-hybrid is not separated during the purification procedure and thus greatly affects the reactivation of papain. This effect decreases the sensitivity of the performed assay and both DNA+ph samples can only be differentiated by a maximal factor of 23.

Further experiments have to be conducted to achieve a reduction in said unspecific interactions of activator and well surface.

## 2.3 Indirect PCR-ALERA

The detailed process of a standard PCR-ELISA procedure was published by Musiani *et al.*[62,63] in 2007. This procedure was taken as a model and the DNA-detection assay developed in this work aims to work analogous as described in the introduction and shown in figure 11.

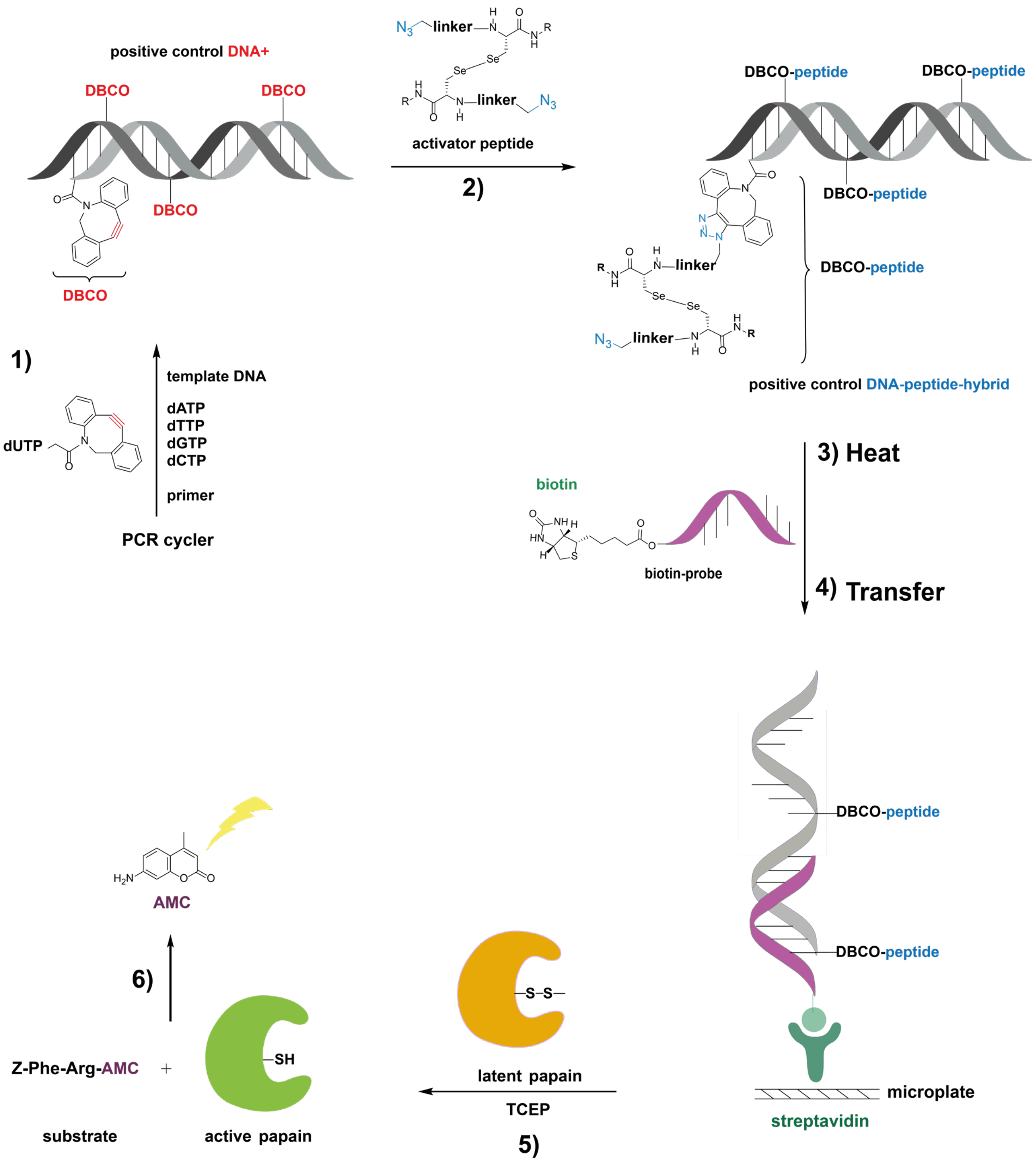

**Figure 11:** Overview of the process of a PCR-ALERA utilizing a biotinylated DNA probe for hybridization to target DNA. 1) PCR with template DNA and modified nucleotide dUTP-DBCO produce a DNA+ amplificate containing DBCO moieties. 2) The PCR mixture is incubated with activator peptide containing an azide moiety for the ring-strain promoted click reaction. 3) The DNA-peptide-hybrid mixture is incubated with a biotinylated DNA probe and hybridized through heating. 4) The reaction mixture is transferred to the streptavidin coated microplate and immobilized while heating. 5) Latent papain, TCEP and substrate are added to the wells. 6) Activated papain processes the substrate and nascent free AMC is detected via fluorescence.

In order to proceed with this last step towards the successful indirect PCR-ALERA that allows for multiplex screening just like the known PCR-ELISA, previous shortcomings have to be overcome. Especially the unspecific binding of activator to the well surface in a yet unknown manner has to be eliminated because results cannot be clearly interpreted otherwise and sensitivity of the assay is reduced greatly. This is still under investigation. If this can be achieved, it is mandatory to verify results and compare them to the commercially available PCR-ELISA assay with regards to specificity and sensibility.

**Acknowledgement**

The authors thank Prof. Dr. Andres Jäschke and Dr. Christian Schmitt for their generous support relating to PCR experiments. This manuscript contains excerpts from the doctoral thesis of author Dr. Miles Bonenberger: M. Bonenberger, “The PCR-ALERA: a PCR-ELISA analogue and antibody-free DNA-detection system”, Dissertation, Universität Heidelberg 2021.